\documentclass[aps,prx,twocolumn,amsmath,amssymb]{revtex4-2}

\usepackage[utf8]{inputenc}
\usepackage{graphicx}
\usepackage{amsmath}
\usepackage{amssymb}
\usepackage{siunitx}
\usepackage{hyperref}
\usepackage{xcolor}
\usepackage{makecell}

\begin{document}

\title{Spectral stability of cavity-enhanced single-photon emitters in silicon}
\author{Johannes Fr\"uh}
\author{Fabian Salamon}
\author{Andreas Gritsch}
\author{Alexander Ulanowski}
\author{Andreas Reiserer}
\email{andreas.reiserer@tum.de}

\affiliation{Technical University of Munich, TUM School of Natural Sciences, Physics Department and Munich Center for Quantum Science and Technology (MCQST), James-Franck-Stra{\ss}e 1, 85748 Garching, Germany}
\affiliation{TUM Center for Quantum Engineering (ZQE), Am Coulombwall 3A, 85748 Garching, Germany}
\affiliation{Max-Planck-Institut f\"ur Quantenoptik, Quantum Networks Group, Hans-Kopfermann-Stra{\ss}e 1, 85748 Garching, Germany}

\begin{abstract} %
The unrivaled maturity of its nanofabrication makes silicon a promising hardware platform for quantum information processing. To this end, efficient single-photon sources and spin-photon interfaces have been implemented by integrating color centers or erbium dopants into nanophotonic resonators. However, the optical emission frequencies in this approach are subject to temporal fluctuations on both long and short timescales, which hinders the development of quantum applications. Here, we investigate this limitation and demonstrate that it can be alleviated by integrating the emitters into Fabry-Perot instead of nanophotonic resonators. Their larger optical mode volume enables both increasing the distance to crystal surfaces and operating at a lower dopant concentration, which reduces implantation-induced crystal damage and interactions between emitters. As a result, we observe a fivefold reduction of the spectral diffusion linewidth down to $\SI{4.0(2)}{\mega\hertz}$. Calculations and experimental investigations of isotopically purified $^{28}\text{Si}$ crystals suggest that the remaining spectral instability is caused by laser-induced electric field fluctuations. In direct comparison with a nanophotonic device, the instability is significantly reduced at the same intracavity power, enabling a tenfold increase of the optical coherence time up to $T_2=\SI{20(1)}{\micro\second}$. These findings represent a key step towards spectrally stable spin-photon interfaces in silicon and their potential applications in quantum networking and distributed quantum information processing.
\end{abstract}   

\maketitle

\section{Introduction}

The implementation of quantum networks~\cite{wehner_quantum_2018} and distributed quantum information processing systems~\cite{main_distributed_2025} requires quantum interconnects~\cite {awschalom_development_2021} with high efficiency that host memory qubits with long coherence time. Different systems are currently explored in this context~\cite{reiserer_colloquium_2022}, ranging from trapped atoms to solid-state platforms, including the nitrogen-vacancy center and other defects in diamond~\cite{ruf_quantum_2021, main_distributed_2025} and rare-earth dopants in other materials~\cite{chen_parallel_2020, kindem_control_2020, ulanowski_spectral_2022, deshmukh_detection_2023, yu_frequency_2023, ourari_indistinguishable_2023, ruskuc_multiplexed_2025}. 

Recently, single-photon emitters and spin-photon interfaces in silicon, including color centers~\cite{redjem_single_2020, hollenbach_wafer-scale_2022, higginbottom_optical_2022, redjem_all-silicon_2023, komza_indistinguishable_2024, day_electrical_2024, simmons_scalable_2024} and erbium dopants~\cite{yin_optical_2013, gritsch_narrow_2022, gritsch_purcell_2023, berkman_observing_2023, berkman_long_2025}, have emerged as a promising alternative that has key advantages for up-scaling: First, the optical transitions of these systems are located in the minimal-loss bands of optical fibers~\cite{holewa_solid-state_2025}, enabling transmission over intercity distances. Second, silicon is an exceptional host for spin qubits that enables hour-long spin coherence times~\cite{saeedi_room-temperature_2013}, as it can be grown at high purity and has a very low nuclear spin density, in particular in isotopically purified material. Finally, nanofabrication has reached a unique level of maturity in silicon, enabling wafer-scale manufacturing of quantum devices~\cite{hollenbach_wafer-scale_2022, rinner_erbium_2023, alexander_manufacturable_2025}.

Despite these advantages, harnessing single emitters in silicon for quantum applications is still in its early stages of development. The key open challenge is to implement a spin-photon interface with high efficiency and spectrally stable emission. Only the former has been achieved based on the Purcell effect in nanophotonic cavities~\cite{gritsch_purcell_2023, johnston_cavity-coupled_2024, islam_cavity-enhanced_2024}, or by collecting the emission via nano-scale silicon waveguides~\cite{gritsch_narrow_2022, komza_indistinguishable_2024, burger_inhibited_2025}. In these devices, however, individual emitters exhibited significant spectral instability: Their spectral diffusion on long timescales so far exceeded $\SI{1}{\giga\hertz}$ for color centers, and $\SI{0.02}{\giga\hertz}$ for erbium dopants, whose 4f electrons exhibit much weaker sensitivity to electric fields. In addition, fluctuations on short timescales severely limited the coherence of the emitted photons~\cite{komza_indistinguishable_2024} and thus the fidelity that can be achieved in remote entangling operations~\cite{photonicinc_distributed_2024}. Even in resonators with strong Purcell enhancement, the dephasing rates have so far significantly exceeded their lifetime-limited values, which substantially hampers quantum information processing.

Recent studies indicate that this fast dephasing is a laser-induced spectral diffusion process~\cite{zhang_laser-induced_2025, bowness_laser-induced_2025}, in which the strong laser pulses exciting the emitters induce charge noise, which alters the optical transition frequencies. The noise is attributed to defects in the silicon lattice, either at device interfaces or introduced into the inner material during device fabrication, particularly during emitter integration. These effects are strongly reduced in bulk crystals, enabling much better spectral stability~\cite{bergeron_silicon-integrated_2020}. However, it is difficult to realize efficient spin-photon interfaces in bulk material, as the Purcell enhancement factor $P$ scales with the inverse volume $V$ of the confined field mode~\cite{reiserer_colloquium_2022}: $P \propto Q/V$, where $Q$ is the resonator quality factor. Thus, Purcell-enhanced emitters in silicon have been studied only in nanostructures so far.

In this work, we overcome the aforementioned challenges by integrating an erbium-doped silicon (Er:Si) membrane of $\SI{2}{\micro\meter}$ thickness into a Fabry-Perot resonator. Compared to nanophotonic devices~\cite{redjem_all-silicon_2023, gritsch_purcell_2023, johnston_cavity-coupled_2024, gritsch_narrow_2022, komza_indistinguishable_2024, burger_inhibited_2025}, the distance of the emitters to the crystal surface is increased at least tenfold, reducing the influence of magnetic fields of impurities and electric fields of patch charges and dangling bonds at the interface. To still achieve a strong Purcell enhancement in spite of the larger mode volume, we use resonators with a high finesse, such that the quality factor of several million exceeds our earlier experiments with nanophotonic erbium devices more than tenfold~\cite{gritsch_purcell_2023, gritsch_optical_2025}, and that with color centers ~\cite{bowness_laser-induced_2025, zhang_laser-induced_2025} by more than two orders of magnitude. With this, we observe a reduction of the emitter lifetime compared to its free-space value, which indicates that the resonator enables efficient single-photon emission into a single mode.

In addition to increasing the distance to interfaces, compared to earlier works we also use a ten- ~\cite {gritsch_optical_2025, burger_inhibited_2025} to hundredfold~\cite{gritsch_purcell_2023} lower dopant concentration of $n_\text{Er}=10^{15}\,\si{\centi\meter^{-3}}$ to achieve a spectral density that allows for individual addressing of single emitters. The reduced concentration has two advantages: First, it reduces the density of defects in the crystal arising from ion implantation, which may cause spectral instability. Second, it reduces interactions between emitters, which deteriorate both optical and spin coherence times with a scaling $\propto n_\text{Er}^{-1}$.

At the used concentration, due to the finite integration yield of less than one percent~\cite{gritsch_narrow_2022}, most nanophotonic devices would not contain a single emitter in their typical wavelength-scale mode volume~\cite{gritsch_purcell_2023}. In contrast, the larger mode volume of the resonators studied in this work allows spectrally multiplexed addressing of hundreds of dopants in the same device~\cite{ulanowski_spectral_2024}. This is enabled by tuning the resonance of the Fabry-Perot cavity on a sub-ms timescale using a piezo tube~\cite{ulanowski_spectral_2022, deshmukh_detection_2023}. 

Based on these advantages, we demonstrate a large improvement in the spectral stability of single emitters in silicon. To gain quantitative insights, we directly compare the results to measurements in nanophotonic devices. In addition, we investigate the remaining sources of spectral instability, both theoretically and experimentally. To this end, we study emitters in a membrane that is isotopically purified in $^{28}\mathrm{Si}$, thereby reducing the nuclear spin density and the associated magnetic field noise by more than a hundredfold.

\section{Sources of spectral instability}

In solids, the optical transitions of embedded emitters are subject to spectral shifts due to their coupling to the solid-state environment, which is mediated by strain, electric, and magnetic fields. In an environment with static disorder, the excited states of different emitters will exhibit random, but static energy shifts. For Er:Si, the resulting distribution of emission frequencies is within $\lesssim \SI{1}{\giga\hertz}$ at several different integration sites~\cite{gritsch_narrow_2022, berkman_observing_2023}. In addition to this inhomogeneous broadening, the optical transitions may be affected by fields that change over timescales that are long or short compared to the lifetime; the former effect is called spectral diffusion, while the latter is referred to as homogeneous broadening.

The three mentioned effects have a different impact on quantum applications. Static disorder in the transition frequency can be counteracted, e.g., by frequency shifting elements~\cite{levonian_optical_2022} or by time-gated photon detection~\cite{vittorini_entanglement_2014}. As it enables spectral multiplexing of up to hundreds of emitters~\cite{ulanowski_spectral_2024}, it can even be an asset in quantum networking~\cite{ruskuc_multiplexed_2025}. In contrast, spectral diffusion and homogeneous broadening are detrimental for quantum applications. While the former can be overcome by tailored decoupling protocols~\cite{ruskuc_multiplexed_2025, uysal_rephasing_2024}, the latter will inevitably reduce the fidelity and/or rate of quantum operations. Therefore, it is paramount to characterize the spectral diffusion and homogeneous broadening of quantum emitters.

\begin{figure*}[ht]    
    \centering
    \includegraphics[width=\textwidth]{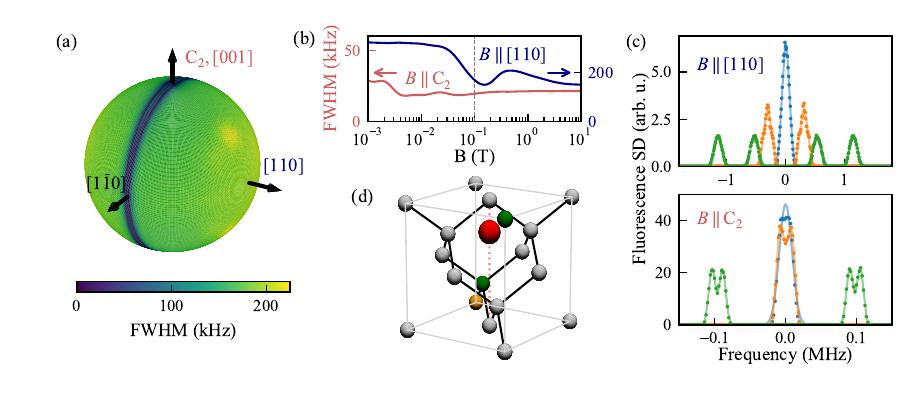}
    \caption{\textbf{Limitation to the spectral stability of erbium caused by nuclear spin noise.} (a) Monte-Carlo simulations are used to calculate the average frequency fluctuations that emitters in site A exhibit owing to the coupling to fluctuating magnetic moments of the nuclear spin bath at the natural isotopic abundance of $^{29}\text{Si}$. The resulting full-width-at-half-maximum (FWHM) of the spectral diffusion linewidth shows a strong dependence on the direction of an applied external magnetic bias field $B$ of $\SI{100}{\milli\tesla}$. (b) The simulation is repeated as a function of the magnetic field in the directions parallel (red, left axis) and perpendicular (dark blue, right axis) to the $\mathrm{C}_2$ symmetry axis of the emitters. (c) At $\SI{100}{\milli\tesla}$, applied perpendicular (top) or parallel (bottom) to the emitter symmetry axis, close-by $^{29}\text{Si}$ nuclear spins in randomly-chosen configurations (colors) can cause a splitting of the lines (orange, green) that depends on their superhyperfine coupling. The individual positions of these strongly-coupled nuclear spins are indicated by the same colors in (d) relative to the erbium (red) position, which is assumed to be displaced by 0.25 unit cells from the unit cell center along the $\mathrm{C}_{2}$ symmetry axis (red dashed line). For the blue curve in (c), no $^{29}\text{Si}$ nuclear spin is found within the unit cell.
    }
    \label{fig:Fig3_Numerical_estimates}  
\end{figure*}

In previous experiments that investigated Yttrium Orthosilicate crystals, the magnetic interaction of individual erbium dopants with the surrounding bath of $^{89}\text{Y}$ nuclear spins (with $100\,\%$ isotopic abundance) has limited the spectral diffusion linewidths to $\SI{0.3}{\mega\hertz}$~\cite{ulanowski_cavity-enhanced_2025}. In silicon, only the isotope $^{29}\text{Si}$ has a nuclear spin. Its nuclear magnetic moment is stronger than that of Y, but it has a much lower isotopic abundance of $4.7\,\%$. Thus, one may assume a similar broadening in both materials. Before turning to an experimental investigation, we first calculate the expected broadening to gain a quantitative understanding.

Specifically, we perform Monte-Carlo simulations in which a single erbium dopant is embedded in a silicon crystal lattice. We then randomly assign $4.7\,\%$ of the silicon atoms to exhibit a randomly oriented $^{29}\text{Si}$ nuclear magnetic moment and calculate the resulting magnetic field at the location of the emitter. With the known $g$-tensors of ground and excited states of site A~\cite{holzapfel_characterization_2025}, the magnetic noise caused by fluctuations of the nuclear spins is translated to a shift of the optical emission frequency. To calculate the resulting spectral diffusion linewidth, we repeat this calculation $3\cdot 10^4$ times, each time assigning random values to the orientation and position of the $^{29}\text{Si}$ spins.

This Monte-Carlo simulation is performed for different orientations of an external bias field of $\SI{100}{\milli\tesla}$; the result is shown in Fig.~\ref{fig:Fig3_Numerical_estimates}(a). Owing to the anisotropy of the g-tensor, the full-width-at-half-maximum (FWHM) of the resulting distribution exhibits a strong dependence on the angle of the magnetic field with respect to the $\text{C}_2$ symmetry axis of the emitters, which is set to the [001] crystal axis. It reaches $\SI{0.2}{\mega\hertz}$ FWHM for most orientations, but can be as narrow as $\SI{0.02}{\mega\hertz}$ FWHM for specific field directions. The uncertainty of these values will be dominated by that of the g-tensor, which is $\SI{6}{\percent}$ for $\text{g}_g$ and $\SI{7}{\percent}$ for $\text{g}_e$~\cite{holzapfel_characterization_2025}.

We repeat the simulation as a function of the external magnetic field at two distinct orientations ${B} \parallel \mathrm{C}_2$ and ${B} \parallel \mathrm{[110]}$, as shown in Fig.~\ref{fig:Fig3_Numerical_estimates}(b). When the external magnetic field exceeds the field generated by the spin bath, it determines the quantization axis of the Er spin, so the optical transition frequency shift is proportional only to the parallel component of the hyperfine interaction, thereby reducing fluctuations. 

The above simulations were averaged over a large ensemble of emitters with different nuclear spin environments. However, in single-emitter experiments, the distribution of $^{29}\text{Si}$ nuclei will be static, and only their orientation will fluctuate, leading to narrower spectral diffusion lines. Remarkably, the magnetic field of close-by nuclear spins can exceed that of the remaining bath, such that one may optically resolve individual nuclei unless other noise sources lead to a larger line broadening. This can be seen in Fig.~\ref{fig:Fig3_Numerical_estimates}(c), which shows the single-emitter spectral diffusion lines expected in three randomly chosen nuclear spin configurations; the location of the $^{29}\text{Si}$ nuclear spins in these configurations is shown in the same colors in panel (d). In the absence of proximal $^{29}\text{Si}$ spins (blue), Gaussian lines of $\SI{146}{\kilo\hertz}$ and $\SI{21}{\kilo\hertz}$ FWHM are observed for the two external magnetic field orientations. The line splits into two (orange) or four (green) in the case of one or two proximal nuclear spins, respectively.

As a result of these simulations, we conclude that superhyperfine interactions are orders of magnitude too weak to explain the observed $\SI{20}{\mega\hertz}$ spectral broadening in previous Er:Si experiments~\cite{burger_inhibited_2025, gritsch_optical_2025}. To verify this experimentally, we will later compare measurements on isotopically purified samples to those in crystals with natural isotopic abundance. Before this, however, we continue the theoretical analysis by calculating the magnetic noise originating from erbium-erbium interactions. To this end, we again perform Monte-Carlo simulations, ignoring the influence of nuclear spins and instead replacing silicon atoms in the lattice by Er, with a probability that matches the experimentally studied concentrations of $10^{15}\,\si{\centi\meter^{-3}}$. We further assume that all erbium dopants exhibit the magnetic moment of site A~\cite{holzapfel_characterization_2025}, which is among the largest of those observed for erbium in any site in Er:Si~\cite{vinh_microscopic_2003, yang_zeeman_2022, berkman_long_2025} and in any other material. Thus, our simulation may be considered an upper bound to the expected spectral diffusion linewidth.

The resulting distribution exhibits an angular dependence that is similar to the broadening caused by the superhyperfine interactions, as shown in Appendix \ref{Appendix_ErErInteractions}. The FWHM ranges from $\SI{0.3}{\kilo\hertz}$ to $\SI{1.3}{\kilo\hertz}$ FWHM, and is expected to increase linearly with dopant concentration~\cite{stoneham_shapes_1969, merkel_dynamical_2021}. It is several orders of magnitude narrower than the broadening expected from the superhyperfine interactions and than the SD linewidths observed in previous experiments~\cite{gritsch_purcell_2023, burger_inhibited_2025, gritsch_optical_2025}.

In addition to the magnetic noise simulated above, spectral instability may also be caused by electric-field noise due to fluctuating charges within the bulk crystal. A numerical estimate would require precise knowledge of the Stark coefficients of Er:Si, which is not currently available. However, the shape of the broadened line and its concentration dependence can be calculated analytically~\cite{stoneham_shapes_1969}. For noise based on randomly fluctuating electric fields of charge traps at a concentration of $\rho$, assuming a linear Stark shift --- as expected from the site symmetry~\cite{holzapfel_characterization_2025} --- one expects a Holtzmark distribution, whose width scales $\propto\rho^{2/3}$. This distribution lies between that of Lorentzian and Gaussian lines. Calculating the exact shape would require additional information about the Stark coefficient. In addition to such random electric fields, the fluctuation of field gradients can also contribute to the SD linewidth. They will lead to Lorentzian lines with a linear scaling $\propto\rho$ with the concentration~\cite{stoneham_shapes_1969}.

In addition to the mentioned sources of spectral instability within the bulk crystal, one may expect significant electric ($E$) and magnetic ($B$) field noise arising from impurities, charge traps, two-level systems, or dangling bonds at the interface. In our experiment, the tenfold increased distance to the surface will reduce field noise more than a hundredfold (as $E_\text{charge}\propto r^{-2}$), and noise related to magnetic $B_\text{dipole}$ and electric $E_\text{dipole}$ dipoles at the interface by more than three orders of magnitude (as $E_\text{dipole}, B_\text{dipole}\propto r^{-3}$). 

Therefore, owing to their different scaling, we can distinguish between noise sources in the bulk and those at the interface by quantifying the SD linewidth of Er:Si in devices with natural and purified isotopic composition, as well as with a tenfold lower Er concentration and a tenfold larger distance to the surface.

\section{Experimental setup}

\begin{figure}[htb]     
    \centering
    \includegraphics[width=\columnwidth]{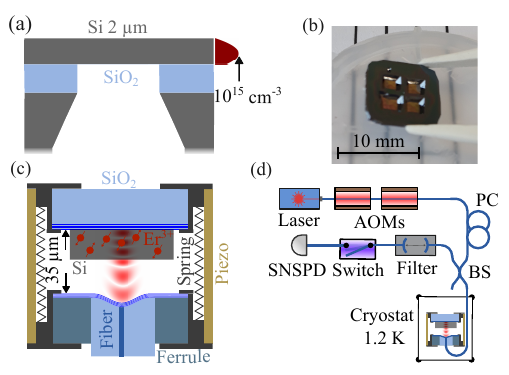}
    \caption{\textbf{Experimental setup.} (a) Sample preparation. The $\SI{2}{\micro\meter}$ thick device layer (grey, top) of a silicon-on-insulator wafer is implanted with erbium dopants (red sketch of the concentration profile), such that a peak concentration of $10^{15}\,\si{\centi\meter^{-3}}$ is obtained. After dicing into chips of $\SI{1}{\centi\meter^2}$, the handle wafer (grey, bottom) and buried oxide (blue) are removed in four quadratic windows of $2\cdot\SI{ 2}{\milli\meter^2}$ using masked chemical etching. (b) Photograph of the silicon chip before transferring one of the four membranes into the resonator. (c) Experimental device (not to scale). The erbium-doped (red spin symbols) silicon membrane (gray) is integrated into the optical mode (red) of a Fabry-Perot resonator. The latter is formed by two distributed Bragg reflectors (dark blue) deposited on a flat mirror substrate (top, light blue) and a concave optical fiber end (bottom), respectively. The mirrors are integrated into a piezo tube (ocher) to allow tuning and stabilization of the resonance frequency. A titanium spring (dark gray) ensures mechanical rigidity. (d) Optical setup. A frequency-stable continuous-wave laser is switched and frequency-shifted by acousto-optical modulators (AOMs). After adjusting the polarization with a polarization controller (PC), it is coupled to the fiber end of the resonator via a 97:3 beam splitter (BS). The reflected light and the dopant emission are guided to a superconducting nanowire single-photon detector (SNSPD) after a fiber-based frequency filter and a switch that prevents detector blinding during excitation pulses. 
    }
    \label{fig:Fig1-setup}
\end{figure}

To experimentally investigate the spectral stability of individual dopants, we enhance their emission by integrating a crystalline membrane into a high-finesse optical Fabry-Perot resonator, as shown in Fig.~\ref{fig:Fig1-setup}. We study two different membranes, both with a thickness of $\SI{2}{\micro\meter}$: The first is obtained from a commercial provider (Norcada) and made from silicon that has been grown by the float-zone (FZ) technique, which is known to provide the highest chemical purity of the grown crystal, thus minimizing paramagnetic impurities. The second membrane uses isotopically purified $^{28}\text{Si}$, grown by chemical vapor deposition on a seed that is formed by a silicon-on-insulator wafer (IceMOS Technology Ltd) with a $\gtrsim\SI{0.03}{\micro\meter}$ device layer using a $>99.9\,\%$ isotopically enriched silane precursor (Lawrence Semiconductor Research Laboratory Inc.).

The crystals are implanted (Helmholtz-Zentrum Dresden-Rossendorf) with erbium to a peak concentration of $10^{15}\,\si{\centi\meter^{-3}}$. This is achieved using an energy of $3.5\,\text{MeV}$ for the FZ and $3\,\text{MeV}$ for the $^{28}\text{Si}$ samples, and a dose of $5\cdot10^{10}\,\si{\centi\meter^{-2}}$. Achieving such a low dose with the used implantation machine requires a very small current, which is difficult to stabilize precisely, leading to significant dose uncertainty and potential concentration differences between the two samples. The realization of high-finesse resonators requires post-implantation annealing at $\SI{600}{\celsius}$ in a rapid thermal annealing oven, with anneal times ranging from $10$ to $\SI{30}{\second}$ and linear ramp-up times of $\SI{60}{\second}$. After this, we fabricate small membranes from the $^{28}\text{Si}$ wafers, see Fig.~\ref{fig:Fig1-setup}(a) and (b). To this end, we dice them after implantation and use masked chemical etching with a $5\,\%$ tetramethylammoniumhydroxide (TMAH) solution for the silicon and hydrofluoric acid (HF) with a concentration of $10\,\%$ for the thermal oxide on the silicon handle wafer and the buried oxide underneath the device layer.

Following their preparation, the membranes are transferred onto a flat dielectric mirror using the techniques developed earlier by our group for the study of erbium in Yttrium-Orthosilicate (Er:YSO) crystals~\cite{merkel_coherent_2020}. Compared to previous experiments with nanophotonic devices of $\SI{0.2}{\micro\meter}$ thickness ~\cite{gritsch_narrow_2022, gritsch_purcell_2023, gritsch_optical_2025, burger_inhibited_2025}, our approach enables investigating single erbium emitters in a tenfold thicker silicon sample, such that the distance to the closest interface is increased accordingly.

A Fabry-Perot resonator is formed by adding a mirror with a Gaussian depression that is fabricated by laser ablation using a setup similar to the one described in~\cite{uphoff_frequency_2015}. Instead of the larger mirror substrates used in our earlier work~\cite{merkel_coherent_2020}, we now achieve a more compact setup by depositing the concave mirror on the end facet of an optical fiber that is glued into a ferrule. This also eliminates the need for free-space coupling, improving robustness. The mirror separation is stabilized and tuned by applying a voltage to a piezo tube, as sketched in Fig.~\ref{fig:Fig1-setup}(c). The lower mass of the smaller mirrors used in this work enhances the agility of frequency tuning by shifting the mechanical resonances to higher frequencies.

After assembly, the devices are fully characterized at room temperature using the methods described in Ref.~\cite{ulanowski_spectral_2022}. The devices are then cooled to cryogenic temperatures in closed-cycle cryostats with home-built vibration isolation systems. For the FZ device, we achieve $\SI{1.2}{\kelvin}$ in a vacuum cryostat (ICEoxford DryIce), while the $^{28}\text{Si}$ devices are investigated in a helium exchange gas at $\SI{1.7}{\kelvin}$ (Attocube AttoDry2100). At both temperatures, the coupling to phonons has a negligible effect on the optical coherence and linewidths of the studied site A in Er:Si~\cite{gritsch_narrow_2022}.

During cooldown, the resonator length is increased by approximately $\SI{10}{\micro\meter}$ owing to the mismatch in the thermal expansion coefficients of the piezo tube and mirror holders. Still, the resonance of the (0,0) transversal electric field mode is tuned to the emission frequency of site A. We then extract the resonator linewidth from reflection measurements and its length from the FSR. We find $\SI{75(9)}{\mega\hertz}$ for the FZ and $\SI{184(21)}{\mega\hertz}$ for the $^{28}\text{Si}$ device, respectively, corresponding to a $Q$ of ${2.6(0.3)}\cdot10^6$ and ${1.06(0.12)}\cdot10^6$. In combination with the radius of curvature and the crystal thickness, which are determined by white light interferometry and reflection measurements, this allows calculating the waist of the resonator mode, which is $\SI{3.8(1)}{\micro\meter}$ and $\SI{2.4(4)}{\micro\meter}$ for the two devices, respectively. With this, the mode volume exceeds that of previously studied nanophotonic resonators~\cite{gritsch_optical_2025} by approximately three orders of magnitude. In direct comparison, the Fabry-Perot cavities thus have a hundredfold smaller Purcell enhancement, despite their tenfold higher quality factors.

For the idealized scenario of a two-level system with a linear dipole at the resonator mode maximum, we would predict Purcell factors of $53(4)$ for the FZ and $63(9)$ for the $^{28}\text{Si}$ device. For erbium, the $\SI{23(5)}{\percent}$ branching ratio of the optical transition to the lowest crystal field level~\cite{gritsch_narrow_2022} reduces this number approximately fourfold; assuming an isotropic dipole moment for the inner 4f shell transitions leads to another 3-fold reduction. With this, we would expect a lifetime reduction with a Purcell factor $P_\text{theo}=4.4$ for emitters located at the maximum of the standing-wave intracavity field.

\section{Inhomogeneous broadening of the emitters}

\begin{figure*}[tb]
    \centering
    \includegraphics[width=\textwidth]{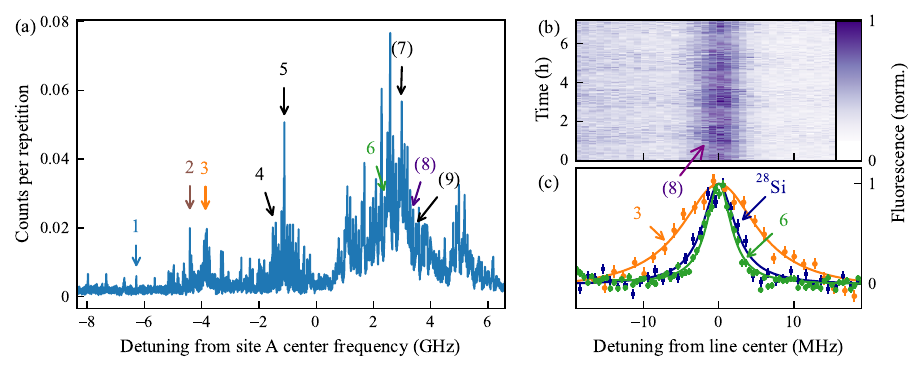}
    \caption{\textbf{Pulsed resonant fluorescence spectroscopy.} (a) When scanning the excitation laser frequency, the fluorescence spectrum contains many sharp peaks, most of which originate from single erbium dopants. We randomly select a set of dopants with strong Purcell enhancement, marked with numbers, for detailed investigation of their spectral stability. At lower power, individual emitters are resolved even in the center of the inhomogeneous distribution. (b) The spectral emission of a randomly chosen emitter, "8", remains stable over many hours. A Lorentzian fit of the time-averaged data exhibits a spectral diffusion linewidth of $\SI{5.2(3)}{\mega\hertz}$ FWHM. (c) In total, we measured 13 emitters with less averaging --- nine on the FZ and four on the $^{28}\text{Si}$ sample. Their SD lines are well-fit by Lorentzian curves (solid lines) and range between $\SI{12.4(12)}{\mega\hertz}$ (orange, "3") and $\SI{4.0(2)}{\mega\hertz}$ (green, "6"). The difference between the emitters shows that the spectral instability is dominated by local fluctuations at each dopant. No linewidth reduction is observed in the isotopically purified sample ("$^{28}\text{Si}$", dark blue, $\SI{5.2(5)}{\mega\hertz}$). High-resolution measurements of the emitters (7), (8), and (9) were performed during a second cooldown.
    }
    \label{fig:Fig2_spectrum_and_inhom_line}
\end{figure*}

In the following, we use pulsed resonant fluorescence spectroscopy to study the properties of individual emitters. Using a fiber-coupled laser setup, shown in Fig.~\ref{fig:Fig1-setup}(d), we first measure a broad spectral overview, as shown in Fig.~\ref{fig:Fig2_spectrum_and_inhom_line}(a). To this end, the resonator is excited by optical pulses of $\SI{0.1}{\micro\second}$ duration, and the fluorescence is then integrated over a time window of $\SI{100}{\micro\second}$ after the pulses, and averaged over $10^4$ repetitions for every detuning of the laser. The latter is defined as the difference between the laser frequency and the center of the inhomogeneous distribution of site A in nanophotonic silicon devices~\cite{gritsch_narrow_2022}.

In this measurement, a high power is used to saturate the transitions far beyond the Fourier-limited excitation bandwidth; thus, increasing the laser frequency in steps of $\SI{2}{\mega\hertz}$ allows for the observation of all dopants in the resonator mode. In parallel with the excitation laser, we adjust the cavity resonance frequency in steps of 100 MHz, which is on the order of the resonator linewidth. This ensures that the emission is collected efficiently while minimizing the time required for discrete mechanical resonator tuning.

In the resulting spectrum, we observe more than 100 sharp lines. They originate from single dopants, which can be shown by measuring the autocorrelation function of the emitted light (see appendix~\ref{appendix_SinleEmitterMM}). In lifetime measurements on some of these peaks, we find values between $\SI{43(2)}{\micro\second}$ and $\SI{82(3)}{\micro\second}$ (see appendix~\ref{appendix_SinleEmitterMM}). Comparing this to the lifetime in bulk, $T_{1,\text{bulk}}=\SI{0.14}{\milli\second}$, we find Purcell factors up to $P=3.3(2)$. The fluctuation between the emitters is caused by the random position of the dopants in the resonator mode~\cite{ulanowski_spectral_2022}. As it is unlikely to find an emitter precisely at the field maximum, the highest value of $P$ is still slightly smaller than the theoretical expectation $P_\text{theo}=4.4$ determined above based on the resonator parameters.

Earlier measurements on erbium ensembles in nanophotonic devices revealed a narrow Lorentzian inhomogeneous distribution of the emission frequency, around $\SI{0.4}{\giga\hertz}$ (FWHM)~\cite{gritsch_narrow_2022}. In contrast, in Fig.~\ref{fig:Fig2_spectrum_and_inhom_line}(a) we find a much broader distribution, spanning more than $\SI{10}{\giga\hertz}$ and grouped into several lines of $\sim\SI{1}{\giga\hertz}$ FWHM. This observation is attributed to strain caused by the difference in thermal expansion between the silicon membrane and the $\text{SiO}_2$ mirror to which it is attached, assuming a strong van der Waals bond is formed. Even if this would create homogeneous uniaxial strain, owing to the low $\text{C}_{2v}$ symmetry of site A~ \cite{holzapfel_characterization_2025}, emitters with different orientations with respect to the lattice would experience different spectral shifts.

A static strain will only broaden the inhomogeneous distribution of the emitters, but is not expected to negatively impact their spectral stability. To investigate this, we perform high-resolution spectroscopy on a randomly-chosen subset of nine emitters in the FZ device with strong Purcell enhancement, marked by arrows in Fig.~\ref{fig:Fig2_spectrum_and_inhom_line}(a), and four emitters in the $^{29}\text{Si}$ device (see appendix). To this end, we first exclude power broadening by reducing the power until the measured linewidths saturate. We then perform a long-term measurement, in which we repeatedly record the spectrum around a single emitter (8) that is well-separated from others. We find that the emission is spectrally stable over many hours and shows no signs of blinking or spectral wandering. A Lorentzian fit to the time-averaged data gives a linewidth of $\SI{5.2(3)}{\mega\hertz}$ FWHM, exceeding the lifetime limit by approximately three orders of magnitude.

The other investigated emitters, both in the FZ and the isotopically purified membranes, show similar values, ranging between $\SI{4.0(2)}{\mega\hertz}$ to $\SI{12.4(12)}{\mega\hertz}$, with an average of $\SI{7.6(23)}{\mega\hertz}$ (see Appendix). The significant variation indicates that the noise that causes spectral broadening is local to each emitter. Comparing the achieved values to measurements in nanophotonic devices at a tenfold larger peak concentration~\cite{gritsch_optical_2025, burger_inhibited_2025}, we observe up to a fivefold improvement.

From these measurements and the mentioned comparison, we can exclude several noise sources as main cause for the observed spectral instability: First, the observed lines are much broader than those expected for Er-Er and $^{29}\text{Si}$-Er interactions, and their scaling does not match that of dipolar interactions with concentration $~\propto \rho$. Second, we can rule out surface noise, as the distance $r$ of the emitters to the closest interface is tenfold increased in our current work, and one would expect improvements $\propto r^{-2}$ for charge noise and $\propto r^{-3}$ for dipolar surface noise when comparing the nanophotonic and FP devices.

Thus, we attribute the residual spectral instability to charge noise from trap states in the bulk material, which scales with concentration. In samples with an average concentration of $2\cdot10^{17}\,\si{\centi\meter^{-3}}$, individual dopants had SD linewidths exceeding $\SI{100}{\mega\hertz}$~\cite{gritsch_purcell_2023}; in samples with $10^{16}\,\si{\centi\meter^{-3}}$, we find $\SI{20}{\mega\hertz}$~\cite{gritsch_optical_2025, burger_inhibited_2025}; in this work, we have  $10^{15}\,\si{\centi\meter^{-3}}$ and linewidths down to $\SI{4}{\mega\hertz}$~\cite{gritsch_purcell_2023}. Thus, we observe a scaling of the SD linewidth of individual erbium dopants in silicon with concentration $n$ that is approximately $\propto{n^{2/3}}$, characteristic of a spectral broadening by random electric fields of point charges~\cite{stoneham_shapes_1969}.

A possible source of these charge fluctuations is found in the erbium dopants themselves. Erbium prefers the triply ionized state in most crystals; nevertheless, earlier measurements in silicon indicated that some of the erbium dopants form shallow donors~\cite{kenyon_erbium_2005}. At cryogenic temperature, these will be in a well-defined charge state. However, they can be ionized or deionized by optical fields, which may then cause the observed spectral diffusion.

\section{Optical coherence}

\begin{figure}[tb]     
    \centering
    \includegraphics[width=\columnwidth]{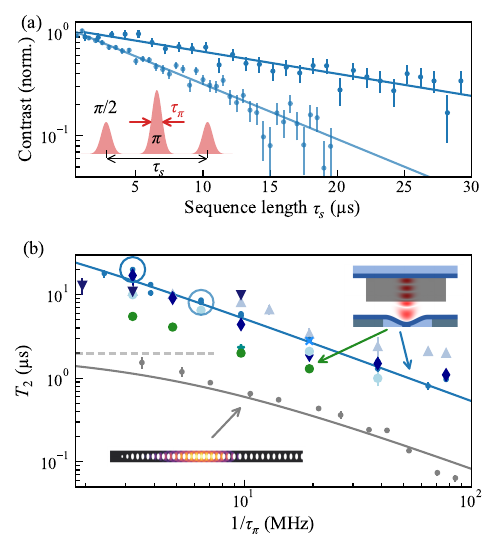}
    \caption{\textbf{Laser-induced decoherence.} (a) (Inset) A photon-echo measurement, consisting of an optical $\pi$ pulse centered in between two $\pi/2$ pulses, allows determining the optical coherence time $T_2$ of individual emitters. The used pulses exhibit Gaussian envelopes with varying pulse lengths $\tau_{\pi}$, while the pulse area is kept constant. (Main panel) Two representative photon-echo measurements on the same emitter, highlighted by blue circles in panel b. (b) The coherence time $T_2$ decreases for $\pi$ pulses of higher intensity and corresponding shorter duration $\tau_{\pi}$, which indicates that laser-induced instantaneous spectral diffusion limits the optical coherence. The data obtained from different emitters in the same device (blue symbols) shows slight fluctuations. A common fit to $T_2=\xi \cdot \tau_{\pi}$ matches the data well (solid lines). A single emitter in an isotopically purified $^{28}\text{Si}$ sample (green) shows the same scaling. A measurement on a nanophotonic device with a tenfold higher dopant concentration (gray) exhibits a much stronger laser-induced spectral instability. Still, it follows the same scaling at short $\tau_{\pi}$, but saturates at long $\tau_{\pi}$, where the curve approaches the lifetime limit (gray dashed line).
    }
    \label{fig:Fig4_photon_echo}
\end{figure}

After studying spectral stability on long timescales, we now turn to the homogeneous linewidth, which describes fluctuations in the emission frequency on timescales shorter than the optical lifetime. To this end, we use fluorescence-detected photon echo measurements on single emitters, as described in detail in Ref.~\cite{ulanowski_spectral_2022}. To this end, we first calibrate the optical pulse area by fitting a sine function to the oscillation obtained in a Rabi measurement with varying amplitude, c.f. our earlier measurements in nanophotonic devices~\cite{gritsch_purcell_2023} and FP resonators~\cite{ulanowski_spectral_2022}. Then, we apply an echo sequence that consists of three optical pulses --- one with a pulse area of $\pi$ that is centered in between two $\pi/2$ pulses, as shown in the inset of Fig.~\ref{fig:Fig4_photon_echo}(a). Subsequently, the delay between the pulses is scanned, and an exponential fit allows for extracting the optical coherence time $T_2$ ~\cite{ulanowski_spectral_2022}.

We find values up to $\SI{20(1)}{\micro\second}$ (dark blue), which decrease when the intensity of the optical pulses $I_\text{Laser}$ is increased (light blue) while reducing their duration $\tau_\pi$ to keep the pulse area constant. In a systematic investigation of this effect, shown in Fig.~\ref{fig:Fig4_photon_echo}(b), we find that the coherence time increases linearly with the pulse duration, $T_2=\xi \cdot \tau_\pi$, for all investigated emitters in the FZ device (blue symbols). The blue least-squares fit includes all data obtained on the FZ sample, with $\xi$ being the only free parameter. While the scaling is the same for all dopants, slight differences in the factor $\xi$ are observed between emitters, which we attribute to fluctuations in the local charge environment. An even larger deviation of $\xi$ is observed for an emitter in the isotopically purified sample (green dots), due to the aforementioned dose- and/or statistical fluctuations.

In our experiment, we vary both the laser intensity $I$ and the pulse duration to produce  $\pi$-pulses of constant area. As the Rabi frequency $\Omega_r\propto \sqrt{I}$, the observed scaling, $T_2\propto\tau_\pi$, corresponds to a laser-induced instantaneous spectral diffusion that is linearly proportional to the absorbed energy $\int_0^{\tau_\pi}I(t)dt$. This is in agreement with recent related measurements of laser-induced decoherence in color centers in silicon, where spectral diffusion increased linearly with laser power at a constant pulse duration\cite{bowness_laser-induced_2025} and linearly with duration at constant power~\cite{zhang_laser-induced_2025}. 

To obtain the longest $T_2$, faint optical pulses would be favored. However, at smaller powers, the condition that the Rabi frequency is much larger than the spectral diffusion linewidth ($\Omega_r \gg \Delta$) is no longer fulfilled in our samples. Thus, the emitters are excited less efficiently, resulting in a reduced fluorescence signal. At the smallest power that still gives enough signal, the homogeneous linewidth of $\Gamma_h=\frac{1}{\pi \cdot T_2}=\SI{16}{\kilo\hertz}$ approaches that of low-power measurements in Er:Si ensembles~\cite{gritsch_narrow_2022}. As the filled outer 5s and 5p shells of rare-earth dopants shield the inner 4f electrons from electric fields, the obtained value outperforms that of T-centers in silicon by more than four orders of magnitude~\cite{zhang_laser-induced_2025, bowness_laser-induced_2025}.

Finally, we investigate if the laser-induced spectral diffusion coefficient $\xi$ depends on the emitter concentration. To this end, we measure a nanophotonic device with a tenfold higher $n_\text{Er}$. The details of this resonator and its fabrication are described in ~\cite{gritsch_optical_2025}. We find significantly shorter optical coherence times, but the same scaling with the laser power at short pulse durations (gray data and fit in Fig.~\ref{fig:Fig4_photon_echo}). In this device, the fit curve saturates for long pulses as the dephasing rate $T_d^{-1}$ of the emitters approaches the lifetime ($T_1$) limit (grey dashed line), such that $T_2^{-1}=T_d^{-1} + \frac{1}{2} T_1^{-1}$.

Comparing the devices, a 6(1)-fold faster laser-induced dephasing is observed at the same laser pulse amplitude. This approximates the $n^{2/3}$ scaling expected for spectral diffusion caused by random electric field fluctuations. To further confirm this, we attempted photon echo measurements in a nanophotonic device with another twentyfold increase in dose~\cite{gritsch_purcell_2023}. However, it turned out that in this case, the coherence was too short for echo measurements.

Based on the above results, we can determine whether the devices can be used for quantum networking via photon interference without compromising the achievable rates~\cite{reiserer_colloquium_2022}. To this end, one has to compare the optical coherence to the lifetime, which is determined by the achieved Purcell enhancement. In the Fabry-Perot devices, $T_2$ falls short of the lifetime limit by a factor of 6.3. Achieving the Fourier limit thus requires either higher Q factors or smaller mode volumes. The latter seems achievable by using mirrors with a smaller radius of curvature~\cite{hunger_laser_2012, bitarafan_-chip_2017, wachter_silicon_2019}. In contrast, $T_2$ approaches the lifetime limit in the nanophotonic resonators. In both devices, further improvement is expected with lower emitter concentration. This, however, would come at the cost of reduced multiplexing capacity unless the yield of erbium integration at site A can be increased substantially.

\section{Summary and outlook}
In summary, by using an optical Fabry-Perot resonator with a $\SI{2}{\micro\meter}$ thick silicon membrane instead of a nanophotonic device, we achieved a tenfold increase in the optical coherence time, and up to a fivefold reduction in the spectral diffusion linewidth of single emitters in silicon. Further improvements in spectral stability may be achieved by using membranes with an even lower dopant concentration in the current device, or by exploring other approaches to efficient quantum interfaces that enable structures with a large mode volume, such as slow-light waveguides~\cite{burger_inhibited_2025}.

The next step towards using such devices for quantum information processing is to implement frequency-multiplexed spin control, similar to earlier works in nanophotonic devices ~\cite{chen_parallel_2020, kindem_control_2020, gritsch_optical_2025}. As our Fabry-Perot resonators achieve a significantly higher $Q$ factor, the spectral selectivity is enhanced. Thus, electron-spin readout can be implemented at much lower magnetic fields $B$, dramatically reducing spin-lattice relaxation via the direct process in which $T_{1,\text{spin}}\propto B^{-5}$~\cite{wolfowicz_quantum_2021}. In addition, Fabry-Perot resonators can target much larger qubit numbers due to two key advantages. First, their larger mode volume enables a reduced dopant concentration, thereby reducing both optical dephasing and spin decoherence through direct interactions that cannot be decoupled efficiently~\cite{merkel_dynamical_2021}. Second, our design enables rapid tuning of the resonator frequency~\cite{casabone_dynamic_2021}, thereby extending the usable spectral region to tens of GHz and covering the entire inhomogeneous distribution. This paves the way to spectrally multiplexed control~\cite{chen_parallel_2020, kindem_control_2020, gritsch_optical_2025} of hundreds of spin qubits~\cite{ulanowski_spectral_2024} in silicon.  

In addition to spectral multiplexing, the presented compact fiber-coupled Fabry-Perot cavity design can enable spatially multiplexed operation of multiple resonators in the same cryostat without requiring free-space optical access. In the future, the Purcell factor of these devices can be further increased by using mirrors with a smaller radius of curvature~\cite{hunger_laser_2012, wachter_silicon_2019}. In addition, our design may be further miniaturized, or direct coupling to on-chip photonic waveguides may be implemented~\cite{bitarafan_-chip_2017, cheng_harnessing_2025}. This would enable much larger qubit numbers than those used in recent pioneering entanglement experiments~\cite{ruskuc_multiplexed_2025}, and thus pave the way for massively multiplexed quantum networking and distributed quantum information processing.

\section{Funding}
Funded by the Deutsche Forschungsgemeinschaft (DFG, German Research Foundation) under the German Universities Excellence Initiative - EXC-2111 - 390814868 and via the individual grant agreement 547245129, and by the European Union (ERC project OpENSpinS, number 101170219). Views and opinions expressed are those of the authors only and do not necessarily reflect those of the European Union or the European Research Council. Neither the European Union nor the granting authority can be held responsible for them.

\section{Acknowledgements}
We thank Flora Segur and Adrian Holzäpfel for their contribution to the numerical spectral-diffusion modeling at an early stage of the project.

\bibliographystyle{apsrev4-2}
\bibliography{bibliography.bib}

\appendix

\section{Expected broadening from Er-Er interactions} \label{Appendix_ErErInteractions}
\begin{figure}[ht]     
    \centering
    \includegraphics[width=\columnwidth]{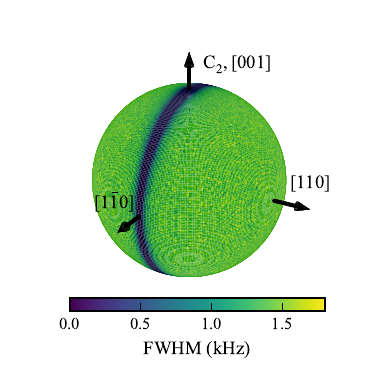}
    \caption{\textbf{Monte-Carlo simulation of the spectral diffusion linewidth caused by Er-Er interactions.} The coupling to surrounding fluctuating erbium dopants at a concentration of $10^{15}\,\si{\centi\meter^{-3}}$ leads to fluctuations of the optical transition frequency, which are much smaller than those caused by the nuclear spin bath; the angular dependence, however, is very similar.
    }
    \label{fig:Fig_SDLinewidth_ErEr}
\end{figure}

To determine the spectral diffusion linewidth expected from Er-Er interactions, we performed Monte-Carlo simulations as described in the main text. The resulting FWHM is shown in Fig.~\ref{fig:Fig_SDLinewidth_ErEr}, exhibiting a similar dependence on the direction of the applied magnetic bias field as the one observed for nuclear-spin-induced spectral diffusion, cf. Fig.~\ref{fig:Fig3_Numerical_estimates}. Owing to the large average distance of $55\,\si{\nano\meter}$ to the nearest neighboring dopant, the FWHM is independent of the magnitude of the magnetic bias field (not shown).

\section{Additional spectra and single-emitter measurements} \label{appendix_SinleEmitterMM}

\begin{figure}[ht]     
    \centering
    \includegraphics[width=\columnwidth]{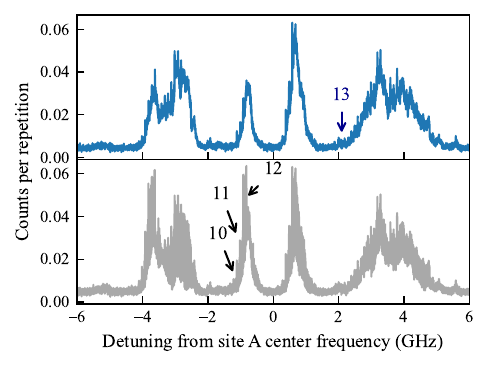}
    \caption{\textbf{Pulsed resonant fluorescence spectroscopy measured using the $^{28}\text{Si}$ device.} Due to a slight ellipticity of the concave cavity mirror, the resonator exhibits a splitting of the fundamental $\mathrm{TEM}_{00}$ polarization modes by a few $\si{\giga\hertz}$~\cite{uphoff_frequency_2015}. Top panel: Measurement of the inhomogeneous distribution using the lower-frequency polarization eigenmode. Bottom panel: Corresponding measurement using the higher-frequency polarization mode.
    }
    \label{fig:Fig_Si28_Spectrum}
\end{figure}

Two different Fabry-Perot resonators were used in this work. The first device contained the FZ membrane whose inhomogeneous distribution is shown in Fig.~\ref{fig:Fig2_spectrum_and_inhom_line}. Two fluorescence spectra of the $^{28}\mathrm{Si}$ device, one for each polarization eigenmode of the resonator, are presented in Fig.~\ref{fig:Fig_Si28_Spectrum}. In both devices, a comparable splitting is observed in the inhomogeneous distribution, attributed to strain arising from the thermal expansion mismatch between the silicon membrane and the underlying mirror substrate.

\begin{figure}[ht]     
    \centering
    \includegraphics[width=\columnwidth]{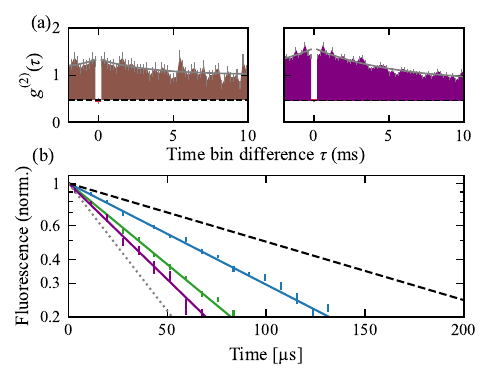}
    \caption{\textbf{Lifetime and autocorrelation measurements.} (a) Autocorrelation function of dopant "2" (left) and "(8)" (right) in Fig. \ref{fig:Fig2_spectrum_and_inhom_line} under pulsed optical excitation. At zero delay, $g^{(2)}(0)$ is in agreement with the value expected for independently measured detector dark counts (black dashed line), proving that the detected photons originate from a single emitter. A bunching behavior is observed within several milliseconds around $g^{(2)}(0)$, as indicated by the exponential fit (grey solid line). It is attributed to spin pumping to a dark state, or vibrations that lead to fluctuations of the resonator frequency. (b) Lifetime measurement of three different dopants, which, due to their random positions in the cavity mode, exhibit varying Purcell enhancement. After subtracting the independently measured dark counts, exponential fits (solid lines) to the data yield lifetimes of $43(2)$ to $\SI{82(3)}{\micro\second}$. Thus, the weakly coupled dopants exhibit a lifetime close to the bulk (black dashed line), while the fastest dopant (purple) approaches the calculated upper bound of the Purcell enhancement (grey dotted line).
    }
    \label{fig:Fig_AutocorrelationMeasurements}
\end{figure}

To confirm that the spectral features in the inhomogeneous distribution arise from single emitters, we analyze the autocorrelation function $g^{(2)}(\tau)$ of the detected photons. To this end, we excite the emitters with short laser pulses of around $\SI{0.1}{\micro\second}$ duration and calibrate the laser power such that the pulse area corresponds to an optical $\pi$-pulse, ensuring efficient excitation. We then measure the autocorrelation function using a single detector and correct it for the independently measured detector dark counts, as described in our earlier experiments~\cite{ulanowski_spectral_2024}. The results are shown in Fig.~\ref{fig:Fig_AutocorrelationMeasurements}(a) for two dopants. The clear suppression of $g^{(2)}(0)$, which approaches zero after dark-count correction, proves that our study is performed on single emitters. 

These emitters are randomly distributed within the cavity mode and thus exhibit different optical lifetimes. This can be seen in Fig.~\ref{fig:Fig_AutocorrelationMeasurements}(b) for three emitters that were randomly chosen out of the subset of dopants that exhibit a large signal in the spectral overview scan and thus a significant Purcell enhancement. It can be seen that the dopant with the shortest lifetime approaches the theoretical limit, calculated using the known branching ratio for site A in Er:Si~\cite{gritsch_narrow_2022} and assuming isotropic optical polarizability for the emitters. 

The clear reduction of the optical lifetime demonstrates that the used structures exhibit a sufficient Purcell enhancement ($P\gtrsim1$) to act as efficient spin-photon interfaces~\cite{reiserer_colloquium_2022}. To this end, it would be beneficial to further increase the emitter excitation probability and photon outcoupling efficiency into the optical fiber forming the resonator, which is currently limited to $5(2)\,\%$. This value is calculated from the detection events per repetition (\SI{0.9(2)}{\percent}) of an emitter far away from the center of the inhomogeneous broadening and the optical losses in the detection path ($\eta_{\text{SNSPD}}=0.82(5)$, $\eta_{\text{Switch}}=0.73(5)$, $\eta_{\text{Filter}}=0.33(4)$, $\eta_{\text{BS}}=0.97(2)$). The coupling efficiency can be improved by integrating gradient-index fibers into the fiber mirror~\cite{gulati_fiber_2017} or by using free-space coupling on the flat-mirror side. In addition, active stabilization using a Pound-Drever-Hall locking scheme is expected to improve the emitter excitation probability, which was not possible in the current configuration because of heating that adversely affected the emitter properties. Instead, a secondary resonator structure may be employed in the future for stabilization, similar to the setup used in our earlier work \cite{ulanowski_spectral_2024}.

In total, we studied 13 dopants in two different samples, as indicated by the numbers in Fig.~\ref{fig:Fig2_spectrum_and_inhom_line} and \ref{fig:Fig_Si28_Spectrum}. In Tab.~\ref{Tab:SingleEmitters}, we summarize the results of these measurements.

\begin{table}  [h!] 
    \begin{tabular}{c|c|c|c|c|c}
        $\#$ & Sample & $P$ & SD LW (MHz) & $\Delta$ (GHz)  & \makecell{Repetitions \\ ($\cdot10^6$)} \\        \hline
        1 & FZ & $1.7(1)$ & $7(1)$ & $-6.29$ & 0.1 \\
        2 & FZ & $2.2(3)$ & $8.0(5)$ & $-4.41$ & 0.5 \\
        3 & FZ & $1.9(1)$ & $12.4(12)$ & $-3.87$ & 0.2 \\
        4 & FZ & $2.3(3)$ & $7.7(24)$ & $-1.45$ & 0.2 \\
        5 & FZ & $2.7(1)$ & $7.4(23)$ & $-1.1$ & 0.1 \\
        6 & FZ & $2.8(5)$ & $4.0(2)$ & $2.43$ & 1.6 \\
        7 & FZ & $2.7(1)$ & $8.2(3)$ & $2.96$ & 0.2 \\
        8 & FZ & $3.3(1)$ & $5.2(3)$ & $3.32$ & 3.1 \\
        9 & FZ & $2.7(1)$ & $5.7(5)$ & $3.42$ &  4.3 \\
        10 & $^{28}\text{Si}$ & $2.1(2)$ & $ 10.2(29)$ & $-1.18$ & 0.1 \\
        11 & $^{28}\text{Si}$ & $2.1(1)$ & $ 10.0(16)$ & $-1.09$ & 0.1 \\
        12 & $^{28}\text{Si}$ & $2.6(5)$ & $ 8.2(21)$ & $-0.88$ & 0.1 \\
        13 & $^{28}\text{Si}$ & $2.6(3)$ & $ 5.2(5)$ & $2.09$ & 0.4 \\

        \hline
        Mean & & 2.4(4) & 7.6(23) &  & \\
        \hline
    \end{tabular}
    \caption{\textbf{Properties of all measured emitters.} As described in the main text, we measured the Purcell factor $P$ and the spectral diffusion linewidth (SD LW) for a total of 13 emitters, characterized by their detuning $\Delta$ from the center of the inhomogeneous line in nanophotonic samples. The last column lists the number of fluorescence spectra acquired for each emitter.}
    \label{Tab:SingleEmitters}
\end{table}

\end{document}